\documentstyle[preprint,aps]{revtex}
\begin{document}
\draft
\title
{Tunneling in the Electron Box in the
Nonperturbative Regime}
\author{G. Falci$^{1,2}$, J. Heins$^1$, Gerd Sch\"on$^1$
and Gergely T. Zimanyi$^3$}
\address{$^{1}$ Institut f\"ur Theoretische Festk\"orperphysik,
Universit\"at Karlsruhe, 76128 Karlsruhe, Germany\\
$^{2}$ Istituto di Fisica, Viale A. Doria 6, 95125 Catania, Italy\\
$^{3}$ Department of Physics, University of California, Davis CA 95616}
\maketitle
\begin{abstract}

We study charging effects and tunneling in the single electron
box. Tunneling mixes different charge states and in the
nonperturbative regime the charge in the island may be strongly
screened. When charge states are nearly degenerate the screening
of the charge is strong even in the weak tunneling regime.
Virtual tunneling processes reduce both the level splitting
$\Delta$ and the tunneling strength $\alpha$.
The charge on the island and the decay rates are calculated.
In the strong tunneling regime also nondegenerate states are
affected by tunneling. Strong-coupling scaling renormalizes the
effective capacitance, a result which we confirm by Monte Carlo
simulations. The tunneling strength $\alpha$ scales to
smaller values into the regime where the weak-coupling scaling
applies. We propose a {\it two stage scaling} procedure
providing the unified picture for the problem. The scaling
analysis is also extended to superconducting tunnel junctions with
finite subgap conductance.
\end{abstract}


\vfill
\eject

\section{Introduction}
Charging effects strongly modify the transport properties of
systems of small tunnel junctions. The transfer of an electron
requires a typical electrostatic energy cost $E_C=e^2/2C$, so
at sufficiently low temperatures ($T \ll E_C)$
charge transfer is suppressed, a phenomenon known as the
Coulomb blockade \cite{kn:AveLik,kn:SchZai,kn:SCT}. Tunneling
induces quantum fluctuations of the charge in the
electrodes. Strong fluctuations screen the charge even at $T=0$
and charging effects are weakened.
Quantum fluctuations are strong if the typical tunneling
resistance $R$ is small enough
($\alpha_0 \equiv \frac{\hbar}{2\pi e^2 R} > 1$).
Even in the weak tunneling regime $\alpha_0 < 1$, quantum
fluctuations are strong if the lowest lying charge
states are nearly degenerate.


A simple and widely studied device where these effects are manifest
is the the single electron box
\cite{kn:saclay}, shown in Fig. 1. It consists of a normal
metal island connected via a capacitor and a tunnel junction to
a voltage source. The energy gap $\Delta$ between the two lowest
lying states can range between $E_C$ and zero and is controlled
by the external voltage (see Fig. 2).
If the tunneling is strong ($\alpha_0 > 1$) the screening of the
charge results in a reduction of the overall effective bandwidth
$E_C$. Also for weaker tunneling, $\alpha_0 < 1$, infrared divergent
tunneling processes strongly mix nearly degenerate charge states.
As a result the gap $\Delta$ is reduced near the degeneracy points,
but the overall bandwidth $E_C$ is unaffected.
The quantitative analysis of these regimes lies beyond the scope
of perturbation theory.
The problem has recently received
much attention \cite{kn:GolZai,kn:Grabert,kn:our_lett,kn:this_vol},
however several previously published results
\cite{kn:GuiSch,kn:Matv,kn:PanZai,kn:Fazio} are in mutual disagreement.
Our motivation is to provide a plausible unified description for this
problem.

In section II we introduce the model. In section III we discuss the scaling
in the weak tunneling regime. In section IV we review the scaling in
the strong tunneling regime and propose a {\it two stage scaling} procedure
to provide a unifying picture for large and small $\alpha_0$.
We also obtained the strong coupling scaling of the effective capacitance
by a Monte Carlo simulation. Then we compare with other non-perturbative
techniques in the $\alpha_0 > 1$ regime. In section V we present results
for other observable quantities, we extend our analysis to finite
temperatures and finally discuss the scaling for a
superconducting junction with subgap quasiparticle tunneling.

\section{The model}
In the absence of tunneling the thermodynamics of the single electron
box is governed by the
electrostatic energy $E_0(Q)=Q^2/2C$ where $C=C_j+C_s$ (see Fig. 1) and
the charge $Q=(Q_x-ne)$ is composed by the $n$ excess electrons in
the island and the continuous ``external charge'' $Q_x=C_sV_x$
induced by the voltage source. The measurable voltage at the junction is
$\langle V\rangle=\langle Q\rangle/C$. The energy spectrum of the
system as a function of $Q_{x}$ is shown in Fig. 2. The lowest lying
levels are degenerate at $Q_x=(k+1/2)e$. At $T=0$ a well defined
$n$ is selected, which depends on $Q_x$ and changes by $\pm 1\,$ when
$Q_x$ crosses the degeneracy points. The result is a step structure
of $\langle n\rangle$ and a sawtooth shape of $ \langle V\rangle$
as a function of $Q_x\;$ \cite{kn:saclay}. At finite temperature both
are smoothened by fluctuations.

In the presence of tunneling the single electron box can be
described by the following Hamiltonian
\begin{equation}
H \,=\, {(Q_x-ne)^2 \over 2C} \,
+\, \sum_{p\sigma} \epsilon(p) \, c^+_{p\sigma} c_{p\sigma} \,
+\, \sum_{k\sigma} \epsilon(k) \, c^+_{k\sigma} c_{k\sigma},\,
+\, \sum_{kp\sigma} T_{kp} \, c^+_{k\sigma} c_{p\sigma} \,+\, h.c.
\quad .
\end{equation}
Here $\,n= \sum_{p\sigma} c^+_{p\sigma} c_{p\sigma}\,$,
$\,c^+_{p\sigma}\,$ and $\,c^+_{k\sigma}\,$ are creation
operators of the electrons in the island and in the lead
respectively, and  $\,\epsilon(p)\,$ and $\,\epsilon(k)\,$
are their kinetic energies.
We consider a wide tunnel junction where tunneling takes place
through $N$ independent channels. In this case the tunneling
strength is given by $\, \alpha_0=N|T|^2 \rho^2\,$,
where $\,T\,$ is the tunneling
amplitude for a single channel and $\rho$ is the densities of states
per channel, which is assumed to be equal in both electrodes.

In order to concentrate on the quantum dynamics of the charge we
integrate out the electronic degrees of freedom \cite{kn:SchZai}
in the partition function $\cal{Z}(Q_x)$.
This is achieved by decoupling the quartic
charging term via a Hubbard-Stratonovich transformation which introduces
the phase $\varphi (\tau)$. After the integration $\cal{Z}(Q_x)$
can be expressed as a path integral depending on the
phase only. Expanding the effective action in powers
of the tunneling amplitudes $T_{kp}$ we obtain a $1/N$ expansion, i.e.
the $2n^{th}$ order term is proportional to
$N |T|^{2n} \rho^{2n} =  \alpha_0^n \,/\, N^{n-1}\,$.
Since in our case the number of channels $N$ is large and the
nominal conductance $\alpha_0$ is finite it is sufficient to retain
only the first term of the expansion, which is given by the
well known path-integral representation \cite{kn:BJMotSch}

\begin{equation}
{\cal Z}(Q_x) \,= \,
\sum_{m=-\infty}^{\infty} e^{2\pi i\, mQ_x/e} \;
\int_{-\infty}^{\infty} d\varphi_0
\int_{\varphi_0}^{\varphi_0 + 2 \pi m} {\cal D} \varphi(\tau) \;
e^{- {\cal S} [\varphi(\tau)]} \quad ,
\end{equation}
with the effective action
\begin{equation}
{\cal S} [\varphi(\tau)] \,=\,  \int_{0}^{\beta} d\tau \;
{1 \over 4 E_C} \left({d\varphi \over d\tau} \right)^2 \;
- \; \int_{0}^{\beta} d\tau \int_{0}^{\beta} d\tau^\prime \;
\alpha(\tau-\tau^\prime) \;
\cos\left[\varphi(\tau)-\varphi(\tau^\prime)\right] \quad .
\end{equation}
The Fourier transform of the dissipative kernel is
$\;\alpha(\omega_{n})=-\pi\alpha_0 |\omega_n|\;$
up to a high energy cutoff $\omega_c$.
The summation over winding numbers,
$\, \varphi (\beta) = \varphi(0) + 2 \pi m \;$, reflects the
discreteness of the charge. The external charge $Q_x$ can be
viewed as a gauge field and appears in the phase factor together
with the winding number $m\,$.

It is useful to consider also the dual representation expressed
in terms of charges
\cite{kn:SchZai}
\begin{equation}
{\cal Z} = \sum_{n=-\infty}^{\infty}
\int_{Q_{x}-n\,e}^{Q_{x}-n\,e} {\cal D}Q(\tau) \,
\sum_{k=0}^{\infty}  {1 \over k!}
\int_{0}^{\beta} d\tau_1 \, ... \, d\tau^\prime_k \,
\alpha(\tau _1-\tau^\prime_1) \, ... \,
\alpha(\tau_k-\tau^\prime_k) \,
e^{-\int_{0}^{\beta}d\tau\,E_0(Q(\tau))} \; .
\end{equation}
Here
$\,Q(\tau) \equiv Q_x - ne -\,e \, \sum_{i}
\Bigl[\Theta(\tau-\tau_i)-\Theta(\tau-\tau_i^\prime)\Bigr]\,$.
Equation (4) maps the tunneling problem on a gas of interacting
blips and antiblips (see Fig. 3) each one representing a tunneling
transition at time $\tau_i$ in which the charge changes
in units of $e$. The tunneling events force a rearrangement of
the other electrons in the two electrodes, which is represented
by the $\,\alpha (\tau)\;$ lines.

We stress that both equations (2-3) and (4) are exact representations
of the original problem in the limit of interest,
$\,N \rightarrow \infty\;$ and $\,\alpha_0\,$ finite.
The extension to finite values of $N$ has also been
considered previously \cite{kn:Matv}.

\section{Weak tunneling regime}
In the weak tunneling regime $\alpha_0 < 1$ it is convenient
to start from the charge representation Eq. (4).
Due to the symmetries of the system we can focus on the interval
$\, 0 < Q_x < e/2 \,$. The main controlling parameters are the
strength of the tunneling $\alpha_0$ and the energy difference
between the two lowest charge states
\begin{equation}
\Delta_0(Q_x)=E_0(Q_x-e)-E_0(Q_x)=E_C(1-2Q_x/e) \quad .
\end{equation}
Due to the infrared divergent behaviour of $\alpha(\omega)$,
perturbation theory in $\alpha_0$ breaks down near the
degeneracy points. Indeed the leading contribution to the
ground state energy, which up to singular terms is
$\,E_G^{(1)}=E_0(Q_x)-\alpha_0\Delta_0\,\ln (\Delta_0/\omega_c)\;$
shows an infrared singularity for $\Delta_0 \rightarrow 0\,$.
Hence both
$\, \langle V \rangle= - {e \over C} \, (d\,E_G^{(1)}/d\,\Delta_0)\;$
and $\, \langle n \rangle=(Q_x-C\langle V \rangle)/e \;$
are logarithmically divergent.

Close to the degeneracy points, $\Delta_0(Q_x)\ll E_C$ the low
energy physics is determined by the lowest two states and involves
tunneling events with energy differences smaller than $E_C$. Thus the cutoff
of the model, $\omega_c$, is approximately given by $E_C$.
In this  two-level-system (TLS) approximation only the terms
$\, n=0,1 \;$ of Eq. (4)
are retained $\,{\cal Z} = \sum_{n=0,1} {\cal G}_n(\beta)\;$
and the only trajectories $Q(\tau)$ contributing
to ${\cal G}_n(\beta)$ are such that blips and antiblips
alternate. For instance we retain the diagrams $(a)$, $(b)$ and $(c)$ in
Fig. 3 and drop $(d)$. The bare propagators are
${\cal G}_n(\tau)= \exp\bigl(-|\tau|\,E_0(Q_x-ne)\bigr)$.

The perturbative corrections are logarithmic
in the regime $\tau_c\ll \tau$.
The first order term (blip-antiblip pair) in the leading logarithmic
approximation reads as
\begin{eqnarray}
{\cal Z}^{(1)}(Q_x,\tau) \,&\approx&\,
e^{-E_0(Q_x)\tau} \, \Bigl( 1 - \alpha_0
(\Delta_0\tau +1)\, \ln(\omega_c\bar\tau) \Bigr) \nonumber\\
&+& \, e^{-(E_0(Q_x)+\Delta_0)\tau} \, \Bigl( 1- \alpha_0
(-\Delta_0\tau +1) \, \ln(\omega_c\bar\tau)\Bigr)
\end{eqnarray}
where $\, 1/\bar\tau=max[1/\tau,\Delta_0,T]\;$
acts as a low frequency cutoff which regularizes the infrared
singularities and $\,{\cal O} (1)\,$ constants depending on the
details of the high frequency cutoff procedure have been ignored.

Our model Eq. (4) is similar to the model used by Anderson et al.
\cite{kn:anderson} to study the single-channel $S=1/2$ Kondo problem.
We could use the same renormalization group (RG) technique to
treat the infrared singularities, namely progressively eliminating
close blip-antiblip pairs by increasing of the short time cutoff
$\tau_c=1/\omega_c$ \cite{kn:our_lett}.
However, since there are technical differences (e.g. here the
interaction is pairwise, and the short time cutoff procedure
is chosen such as to guarantee that $\alpha(\omega_n=0)=0$),
some care is needed when using the scaling technique of
Ref. \cite{kn:anderson}.
We chose to eliminate high frequencies; so we split the kernel
$\alpha(\tau)$ into slow and fast parts, the latter containing
the frequencies we want to integrate out,
$\, \alpha(\tau)=\alpha_s(\tau)+\alpha_f(\tau)\,$.
In first order in the tunneling strength we look at all the
configurations containing only one ``fast'' $\alpha$-line which
can connect either a close blip-antiblip pair or a pair
separated by some blip-antiblip insertion.
Only the former are effective in the first step of the
renormalization. Indeed the integration of the {\it complete}
(fast {\it and} slow modes) line of the latter kind does not give
infrared singularities because of phase space restrictions,
thus we discard them.
Interaction lines overarching blip-antiblip insertions in
higher order diagrams (eg. the ``rainbow'' of Fig. 3b) will enter
the renormalization in the successive steps, once the blip-antiblip
insertions are eliminated.
In this procedure the running parameters will never be
renormalized by diagrams containing crossing $\alpha$-lines. Our
treatment is equivalent to the ``Non-Crossing Approximation''
and can be justified by direct calculation. Indeed elimination a
close pair with crossing $\alpha$-lines (see Fig. 3c) yields an
interaction proportional to $\tau^{-3}$ which does not lead to
any logarithmic singularity.
The bare ground state propagator ${\cal G}_0(\tau)$ is renormalized
in lowest order by the fast modes
$\, \omega_c-\delta\omega_c < \omega < \omega_c \;$
of a single blip-antiblip close pair which generate a contribution
$\, -\alpha_0 (1+\Delta_0 \tau) (\delta\omega_c/ \omega_c)\,$.
A similar expression gives the renormalization of
the propagator for the first excited state ${\cal G}_{1}(\tau)$.

The general $n$-th order terms can be cast in a way such that the
partition function preserves the original form if the gap and the
interaction scale as
\begin{equation}
{d\,(\Delta/\omega_c)\over d\, \ln \,\omega_c}\,=\,
\Bigl( 2\alpha Z^2\,-\,1 \Bigr) {\Delta \over \omega_c} \; ,
\quad
{d\,(2\alpha Z^2) \over d\, \ln \,\omega_c} \,=\,
\Bigl( 2\alpha Z^2 \Bigr)^2  \quad .
\end{equation}
The renormalized ground state energy is given by
$\, E_G = E_0(Q_x)+ 1/2 (\Delta_0 - \Delta) \,$.
Here $Z$ is the wave function renormalization which enters always
together with the interaction in the combination $\, 2\alpha Z^2 \,$,
and the $-1$ in the first equation enters because of the dimensional
factor $\omega_c \,$.

The scaling equations can be readily integrated down to a
low energy scale $\omega_c$
\begin{equation}
\alpha(\omega_c) \,=\, \alpha_0 \,\, \Bigl( 1 + 2 \alpha_0 \,
\ln \, ({E_C \over \omega_c}) \Bigr)^{-1}  \; ,
\quad
\Delta(\omega_c) \,=\, \Delta_0 \,\, \Bigl( 1 + 2 \alpha_0 \,
\ln \, ({E_C \over \omega_c}) \Bigr)^{-1} \quad .
\end{equation}

At zero temperature the renormalized gap provides the low energy
cutoff so we have to stop the scaling at $\omega_c\,=\,\Delta$.
Then Eq. (8) becomes a self-consistent equation for $\Delta$.
At finite temperatures, if $T > \Delta$ the RG has to be stopped
at $\omega_c= max[\Delta,T] \,$, where the infrared singularities
in the perturbation expansion disappear.

Typical solutions
are shown in Fig. 4. Notice that the renormalized gap $\Delta$ does
not vanish.

\section{Strong coupling regime}
In the regime $\alpha_0 \gg 1$ the charge fluctuates strongly
so it may be convenient to start from the phase representation
Eqs. (2-3). We analyze the problem by various nonperturbative techniques.
We start with the RG analysis of Refs.
\cite{kn:GuiSch,kn:Kosterlitz,kn:bulgadaev}.
This treatment is restricted to the $m=0$ sector of the partition
function Eq. (2), so the discreteness of the charge is not
explicitly accounted for. However, we will show that the overall
conclusions of the scaling theory can be carried over to the general
case, described by Eq. (2).

The scaling equations are obtained perturbatively in $1/\alpha \,$
\cite{kn:GuiSch,kn:Kosterlitz,kn:bulgadaev}.
\begin{equation}
{d\,(E_C/\omega_{c})\over d\, \ln\,\omega_{c}}\,=\,
\Bigl({1\over\tilde{\alpha}}\,-\,1\Bigr)\, (E_C/\omega_{c})\, ;
\quad
{d\,(1/\tilde{\alpha})\over d\, \ln\,\omega_{c}}\,=
\, -1/\tilde{\alpha}^2 \quad ,
\end{equation}
where $\tilde{\alpha} = 2\pi^2\alpha$.
In the present approach we determine
the scaling of the overall bandwidth $E_C$, but not that
of the gap $\Delta$, exactly the opposite of the weak coupling region.
However we assume that the gap $\Delta$ scales similarly to the
bandwidth. Equation (9) shows that $\tilde \alpha$ decays in the
strong coupling regime towards weak coupling; so in the case of
strong tunneling a {\it two step scaling procedure is called for}
which provides the desired unified
picture. As the energy scale $\omega_c$ (e.g. temperature)
decreases, first $\alpha$ decays from its large initial value
${\tilde\alpha_0}$ according to Eq.(9)
\begin{equation}
{1 \over \alpha\,(\omega_c)} \,=\,
{1 / \tilde{\alpha_0} \over
1 - (1 / \tilde{\alpha_0}) \, \ln\,(\omega_{c0}/\omega_c))}
\end{equation}
to $\, \tilde\alpha \sim 1$. The bandwidth also decreases upon
renormalization and for $\tilde\alpha \sim 1$ reaches the value
$\, E_C^*\,\approx\,E_C^0\, \exp\,(-\tilde{\alpha_0}) \,$. At this
point the cutoff reaches $\omega_c\sim E_C^*$.
{}From there on we use the weak coupling scaling of section III with
initial values $\omega_c\approx E_C^*\,$, the exponentially
suppressed bandwidth,
$\,\Delta_0\,\approx\,\Delta_0\, \exp(-\tilde{\alpha_{0}})\,$, and
$\tilde\alpha_0\approx 1$ implying $\alpha_0\approx 1/2\pi^2$.
The flow of $\alpha\,(\omega_c)$ will be governed
by Eq.(7). Again the tunneling strength decreases and eventually we
stop the RG when we reach the low-energy cutoff. The final formula for
the gap in the $\alpha_0 \gg 1$ regime is
\begin{equation}
\Delta(\omega_c) \,=\,
{ \Delta_0 \, e^{-\tilde{\alpha_0}} \over
1+ \pi^{-2} \, \ln \, (E_C^0 \, e^{-\tilde{\alpha_0}} / \omega_c)}
\quad .
\end{equation}

Next we checked the exponential suppression of the bandwidth by a
Monte Carlo simulation starting from the phase representation
in the normalized form
\begin{equation}
{\cal Z}(Q_x) \,=\, {1 \over \sum_{m} {\cal Z}(m)}  \,\,
\sum_{m=-\infty}^{\infty} \,\, {\cal Z}(m) \; e^{2\pi i m Q_x/e}
\quad ; \quad
{\cal Z}(m) \,=\, \int{\cal D} \, \delta\varphi(\tau) \;
e^{- {\cal S}_m[\delta\varphi]}
\end{equation}
where we decomposed
$\varphi(\tau)=\delta\varphi(\tau)+\varphi_0+2\pi m \tau/\beta\,$,
$\;{\cal S}_m\,[\delta\varphi(\tau)] \, \equiv \,
{\cal S}[\delta\varphi(\tau)+2\pi m \tau/\beta]\;$.
and the fluctuations satisfy
$\, \delta\varphi(0)=\delta\varphi(\beta)=0 \,$.
At this stage we mention some technical points
\cite{kn:Simanek,kn:Scalia}. First, the part of the action which
describes tunneling is equivalent to a one-dimensional classical
XY model with long range interaction where
$\beta E_C^0 \,$ plays the role of the system size.
Then each update of the
phase $\delta\varphi(\tau_i)$ requires a summation over the whole
discretized lattice. We update $\delta\varphi(\tau_i)$ using
the scheme proposed in Ref. \cite{kn:Simanek} but in a standard
Metropolis algorithm. Second, in order to extract
$E_C^{*}$ we need
$\; 1 \ll \beta E_C^{*}\sim \beta E_C^0\, e^{-\tilde{\alpha_0}} \,$.
Hence large values of $\tilde\alpha_0$ require large $\beta E_C^0 \,$
i.e. a large number of lattice sites. Third, it is apparent from
Eq. (12) that ${\cal Z}(m)$ is proportional to the probability for
the path to visit the $m$-th sector. Therefore we cannot perform
separate simulations for each sector (we would trivially obtain
${\cal Z}(m)=1 \,$).
On the other hand simulating continuous and discrete variables
altogether is a technically difficult task.
The most economical scheme we found is to calculate
\begin{equation}
{{\cal Z}(m) \over {\cal Z}(m+1)} \,=\,
{\int{\cal D} \,\delta\varphi \;
e^{-{\cal S}_m}\,\,\,e^{{\cal S}_m - {\cal S}_{m+1}}
\over \int{\cal D}\,\delta\varphi \; e^{-{\cal S}_m}}
\end{equation}
sector by sector. Notice that we are interested to the marginal
probability for a jump to a different $m$ sector whereas in the
Monte Carlo procedure we update
$(\delta \varphi,m) \rightarrow (\delta \varphi^{\prime},m^{\prime})$.
This move can be carried out in several ways, according to how we
choose the final $\delta \varphi^{\prime}$ configuration. Also
the relaxation of the path $\delta \varphi$ can take place by
updating sequentially or randomly in the lattice. The results
we present do not depend on the details described above. In each
simulation we measure 300 sample points per winding number.
Between two sampling points the system had time to evolve for 7
more sweeps. We measured the Gibbs energy at lower and lower
temperatures, i.e. increasing the number of lattice sites for
the path $\delta \varphi$ up to 1000 at the lowest temperatures.
One can obtain the normalized
${\cal Z}(m)$ from the results Eq. (13) by recursion relations
and extract the ground state energy $E_G(Q_x)$ from finite size
scaling. Further details are given in \cite{kn:Jochen}.

Reliable results are obtained close to $Q_x \approx 0 \,$,
and the effective bandwidth (inverse capacitance)
has been extracted from the
curvature of the band (see Figs. 4 and 5).
However the region close to the band edges is beyond the
capability of the Monte Carlo method. Indeed  ${\cal O} (1)$ terms
in the winding number summation Eq.(12)
enter with the oscillating phase factor
$\, \exp\,(2\pi i m Q_x/e) \,$.
Close to the bottom of the
band $Q_x \approx 0\,$, and the oscillations are slow, whereas
for $Q_{x}\approx e/2 \,$ they are fast. At the band edges
one expects
${\cal Z}(Q_x\sim e/2) \sim \exp\,(-\beta\,(E_C-\Delta))\ll 1 \,$.
Hence one has to extract an extremely small number
by adding many ${\cal O} (1)$ terms with oscillating phase
factors. This makes the numerical procedure very unstable near
$\, Q_{x}= e/2 \,$.

We stress that the same argument
applies to any analytic calculation which attempts to estimate
${\cal Z}(m)$ and perform the winding number summation.
Examples will be discussed below.

Panyukov and Zaikin\cite{kn:PanZai} studied the strong coupling
regime by a non-standard instanton technique in the phase representation
Eqs. (2-3).
They obtained a renormalized bandwidth
$E_C^{*}\,\sim \,E_C\, \tilde{\alpha}^{2}\,\exp\,(-\tilde{\alpha})\,$,
which agrees with the results of large $\alpha$ scaling,
as well as the Monte Carlo data down to surprisingly
low values of $\alpha_0 \,$. The result is shown in Fig. 5
where the curve of Panyukov et al. has been rescaled by a
factor $\sim 2$, i.e. also the accuracy in the pre-exponent
is remarkable.

Panyukov and Zaikin further concluded that the ground state
energy {\it completely flattens} in a wide interval around
$Q_x = \pm e/2 \,$. This latter conclusion differs from
our picture and below we will argue that the possible reason
for this discrepancy is the lack of accuracy of the instanton
calculation which becomes crucial near the degeneracy points.

The technique presented in \cite{kn:PanZai} is unusual
in two senses: i) each instanton possesses the standard
zero mode related to the location, but also a non-standard one
related to their width. Thus instantons of all lengths
enter ${\cal Z}(Q_x)$ with similar weight.
ii) A summation over winding numbers $m \,$ connected with
oscillating phase factors has to be performed (cf. Eq. (12)).
Hence, near the band edges innocent
looking approximations can profoundly alter the result.
We reinvestigated the accuracy and consistency of the method used in
Ref.\cite{kn:PanZai}. In particular a dilute instanton
approximation has been used which in standard fixed size instantons
calculations is justified when the $\cal N$-instanton
configurations playing the main role have finite $\cal N \,$.
Then for fixed size instantons the dilute limit is always
reached for $\beta \rightarrow \infty \,$.
As the size of the instantons is not limited here, we have to
calculate the average width
$\langle \sigma \rangle_{\cal N} \;$ of an instanton in an
$\cal N$-instanton configuration and the dilute limit
is reached only if
$\cal N \langle \sigma \rangle_{\cal N} \ll \beta \;$
for the relevant values of $\cal N \,$.
In the partition function for $\cal N$-instantons
$\, {\cal Z}(\cal N) \,$
we first separate the zero modes (the locations $\tau_j$ and the width
$\sigma_j$ of each instanton) and integrate the remaining fluctuation
determinant to get
\begin{equation}
\langle \sigma_i\rangle_{\cal N} \,=\,
{{\cal N}! \, \Bigl( 4T\,E_C^* \cos\,(2\pi Q_x/e)\Bigr)^{\cal N}
\over {\cal Z}(\cal N)} \;
\int \prod_{j}^{\cal N} \,d\tau_j \,\,
\int \prod_{j}^{\cal N}\,d\sigma_j \;  \sigma_{i} \;
=\,\, {\beta \over 2{\cal N}+1}  \quad .
\end{equation}
As in Ref. \cite{kn:PanZai} we constrain the integration
to non-overlapping instanton configurations
($\, \tau_j+\sigma_j/2 < \tau_{j+1}-\sigma_{j+1}/2 \,$).
Hence the result (14) is a {\it lower} limit to the average
size and $\cal N \langle \sigma \rangle_{\cal N} \,$ is never
small compared to $\, \beta \,$. We conclude that the
instanton gas is {\it never}
dilute. Thus the contributions from overlapping instanton
configurations are comparable to the non-overlapping ones
considered in \cite{kn:PanZai}.

We calculated also the
bare interaction between an instanton and an anti-instanton
$\; {\cal S}(i,j) \,=\, -\,16\tilde\alpha \sigma_i\sigma_j/
[(\sigma_i+\sigma_j)^2 \,+\,
(\tau_{i}-\tau_{j})^2] \;$ which then attract each other.
Instantons of the same ``sign'' do not interact.
Close pairs of instantons are favoured, and this will amplify
the deviations from the dilute, non-interacting instanton gas
picture. Since close to the degeneracy points
extreme accuracy is required we consider the result at the edges
of the bands derived in
Ref. \cite{kn:PanZai}, which are based on considerable
approximations, not conclusive.
However close to the bottom of the band results are less
sensitive to approximations, which explains the excellent
agreement between the instanton
calculation and our Monte Carlo results.

The exponential suppression of the bandwidth is also found in
the straight-line approximation, introduced in \cite{kn:Fazio}.
We make use of the decomposition of $\varphi(\tau)$, introduced
for the Monte Carlo simulation, and expand the action in terms
of the fluctuations $\delta\varphi(\tau)$ around the straight
lines $2\pi m \tau/\beta+\varphi_0$
\begin{equation}
{\cal S}_{m}(\delta\varphi)\,\approx\, \tilde\alpha\,|m| +
T\sum_{\nu>0}\,\Bigl[{\omega_{\nu}^{2}\over 4\,E_C}\,+\,
{\tilde{\alpha}\over2\pi}\, (|\omega_{\nu}
-\omega_{m}|\,+\,|\omega_{\nu}+\omega_{m}|\,-\,2|\omega_{m}|)\Bigr]\,
|\delta\varphi_{\nu}|^2 \,\, .
\end{equation}
This action gives rise to soft modes for $|\omega_{\nu}|<|\omega_{m}|$.
The fluctuation determinant relative to the $m=0$ term is
\begin{equation}
{det(m)\over det(0)}\,=\, a^{-m}m!\, \prod_{\nu=1}^{m} {1\over 1+\nu/a}
\,\prod_{\nu=m+1}^{\infty}\bigl[1-{m\over\nu(1+\nu/a)}\bigr]
\end{equation}
where $a \equiv {2E_C \tilde \alpha \over \pi^2 T}$. In the $a \gg 1$
limit, we obtain the partition function
\begin{equation}
{\cal Z}(Q_{x})\,\approx\,\sum_{m\ll a} {a^{m}\over m!}
e^{-\tilde{\alpha}\,|m|} e^{i2\pi mQ_x/e}
\end{equation}
In the temperature regime $E_C\tilde{\alpha} e^{-\tilde{\alpha}}\ll T\ll E_C$
we can confine ourselves to $m=0,\pm 1$. This yields
$(1/\beta)\,\ln{\cal Z}(Q_x)\approx const + E_C^*\, \cos\,(2\pi Q_x/e)$ where
$E_C^{*}\,=\,{4E_C\tilde{\alpha}\over\pi^2}\, e^{-\tilde{\alpha}}$
in agreement with the previous results. However at lower temperatures
the approximation breaks down and the numerical summation over $m$ gives
a negative partition function. This breakdown is understandable since
it happens at the crossover temperature where according to the scaling
analysis the effective $\alpha$ decreases below 1, and quantum fluctuations
of $\delta\varphi(\tau)$ become large.

\section{Other results and discussion}
A scaling analysis, very similar to the one we present in section III,
was performed in Ref.\cite{kn:GuiSch}.
The gap renormalization of Eq. (7a) had been found there, but the
possibility of wave function or $\alpha$ renormalization was not
considered. As a result a phase transition between a finite gap and
a zero gap region at $\alpha_0=1/2$ was predicted.
The main consequence of the additional scaling of $\alpha$ of Eq.(7b)
is that the gap remains finite. The transition is smeared, leaving only a
strong crossover around $\alpha_0 \sim 1/2$ (see Fig. 4).

Two studies addressed directly the weak coupling regime,
one performing a poor man's scaling analysis\cite{kn:Matv}, and
one solving a Dyson equation\cite{kn:GolZai}.
Our results agree with those in the leading logarithmic approximation.
Difference arises because
in our formalism the gap appears as an explicit low energy cutoff,
yielding a self-consistent equation for $\Delta$.
This will be important when $\alpha_0$ is not very small.

The ground state energy $E_G$, the voltage at the junction
$\;C \langle V \rangle \,= \,d\,E_G(Q_x)/d\,Q_x\;$
and the average number $\;\langle n \rangle\,$ of electrons
on the island are shown in Figs. 6, 7 and 8.
As Ref.\cite{kn:PanZai} found a complete flattening of the band
around the degeneracy points, it was suggested that in the middle of the
vertical part of the original sawtooth pattern of $\langle V \rangle$
\cite{kn:sawtooth} a new $S$-shape develops.
In the light of the above analysis there is no support for the complete
flattening. As no reliable treatment of the gap is available in the large
$\alpha$ regime, we used the scaling formalism in the $\alpha_0\sim 1$ regime
(see Fig. 7). The logarithmic corrections
of Eq.(7) modify the vertical part in a weak manner. The clearest
consequence is the strong suppression of the amplitude of the sawtooth
oscillations already at moderate values of $\alpha_0$.

Another observable feature is the possible broadening of the excited states,
caused by the appearance of a finite lifetime $\Gamma^{-1}$.
We adopt the method of Ref.\cite{kn:GolZai} to determine $\Gamma$
close to the band edges, in the $\alpha_0 < 1$ regime.
Recall that for $Q_x < e/2$ the ground state energy close to the
degeneracy point is given by
$E_G(Q_x) \,\approx\, E_C/4 - \Delta (Q_x)/2\,$.
We can determine the energy of the first excited state $E_1(Q_x)$
by analytic continuation of $E_G(Q_x) \,$ to $\, Q_x > e/2 \,$.
The argument of the logarithm in $\Delta$, turns negative when
passing $Q_x=e/2$ and an imaginary part develops in $E_1(Q_x)$.
The inverse lifetime is then determined by
\begin{equation}
\Gamma \,=\, \Im m \, \Bigl( E_1(Q_x) \Bigr) \,\approx\,
{4 \pi \alpha_0 \, \Delta_0 \over
\Bigl( 1 + 2 \alpha_0 \, \ln(E_C / \Delta_0) \Bigr)^2
\,+\, (4 \pi \alpha_0)^2}      \quad .
\end{equation}
As $\Gamma \ll \Delta$,
the excited levels remain well defined \cite{kn:schoeller}.
The primary experimental consequence of $\Gamma > 0$ is that
the I-V curves become smoothed proportional to $\Gamma/\Delta$
around the onset of the Coulomb-blockade.

At finite temperatures also thermal fluctuations have to be considered.
If $T \ll E_C$ and for $\alpha_0 < 1$ the physics near the band edges
still involves only the two lowest charge states and the Gibbs
energy is given by
\begin{equation}
G(T,Q_x) \approx E_0(Q_x)+ \Delta_0/2 - {1 \over \beta}\,
\ln\, \bigl[\, 2\, \cosh \, \bigl({\beta\Delta \over 2} \bigr) \bigr]
\end{equation}
The renormalized gap is then calculated using Eq.(8) with
$\,\omega_c = T\;$ for $\,T>\Delta(Q_x)\;$ and with
$\,\omega_c = \Delta\;$ for $\,T<\Delta(Q_x)\,$. All the quantities of
interest can be calculated and in particular the normalized
$\langle V \rangle$ vs. $Q_x$ has now a finite slope at $Q_x=e/2\,$,
given by $\,\,({1\over 2} \beta E_C)/[1+2\alpha_0 \, \ln(\beta E_C)]^2\,$.
This result has also been found by other methods \cite{kn:schoeller}.
Notice that a finite tunneling strength is very effective in
suppressing the slope when $\,2\,\alpha_0 \, \ln(\beta E_C) \sim 1\,$.
In the experiments of the Saclay group \cite{kn:saclay},
$\;2\alpha_0 \, \ln(\beta E_C) \sim 10^{-2}\;$, so quantum fluctuations
of the charge do not explain the observed suppression of the slope
at $Q_x=e/2$. In this case the screening of
the charge is probably due to the fact that thermal noise coming
from the electromagnetic environment can excite tunneling\cite{kn:FaBuSch}.

Finally we reconsider the case of S-S junctions with finite
subgap quasiparticle tunneling.
Here the effect of a Josephson coupling between the electrodes has to be
considered as well. For large $E_J$ a Kosterlitz-Thouless type
transition was found in Ref.\cite{kn:GuiSch}: for $\alpha_0>1/4\;$ the
Josephson coupling $E_J$ scales towards larger values.
Here we discuss the small $E_J$ limit, where $E_J$ does not
renormalize (see below).
The flow diagram in the $E_J/E_C$--$\alpha$ is shown in Fig. 9.
plane. In the strong tunneling regime the flow lines of are
given by $E_J/E_C \propto \exp(- \tilde \alpha_0)\;$ \cite{kn:GuiSch}
which is due to the exponential suppression of $E_C$ discussed in
section IV. One of these lines is the separatrix between the two
phases which ends at $\alpha = 1/4$ for $E_J \rightarrow \infty$.
In the regime $\alpha < 1$ (i.e. $\tilde \alpha < 2 \pi^2$) we can
study at the charge representation Eq. (4), modified by the effect of
the Josephson tunneling \cite{kn:SchZai,kn:GuiSch}. The typical
configurations differ from those shown in Fig. 3 because $2e$ blips
and antiblips are present, due to the transfer of Cooper pairs.
However, {\it no infrared process is connected with the transfer of Cooper
pairs and hence $E_J$ does not renormalize}.

We thus arrive at the following picture for the the small $E_J$ regime:
for large $\alpha$
the system flows towards larger $E_J/E_C$ until it reaches
$\tilde \alpha < 1$ ($\alpha < 1/2 \pi^2$) where $E_C$ itself does
not scale appreciably anymore, and the flow lines flatten.
In general $\alpha$ will scale towards smaller (cf. Eq.(8)) but
finite values, as discussed in section III. Thus the flow lines in
Fig. 9 do not reach the $\alpha=0$ axis.  The actual final value
of $\alpha$ depends on the single particle gap $\Delta_0(Q_x)$.
Also the detailed behaviour of the separatrix in the intermediate
$\alpha$ and $E_J$ regimes may depend on $\Delta_0$.


We acknowledge discussions with E. Brezin, R. Fazio,
D. Esteve, R. Fresard, H. Grabert, F. Guinea, H. Pothier, H. Schoeller
and A.D. Zaikin.
This work is part of the Sonderforschungsbereich 195, supported by the DFG.
GTZ has been supported by NSF grant 92-06023. GF has been supported by EC
grant ERBCHBICT930561 and by Della Riccia Foundation.

\vspace{25cm}

\begin{large}
\noindent
Figure Captions :
\end{large}
\vspace{0.8cm}

\noindent
Fig. 1 : The single electron box. It shows
the interplay between the discrete charge $\,-ne \,\,$ on the
island and the continuous charge $\, Q_x=C_sV_x \,$ controlled by the
voltage source.
\vspace{0.5cm}

\noindent
Fig. 2 : The band structure for $\alpha_0=0$. The energy gap $\Delta_0$
can be tuned between $E_C$ and zero by varying the external voltage.
\vspace{0.5cm}

\noindent
Fig. 3 : Some diagrams contributing to $\cal Z$. Full lines
represent the allowed trajectories in the charge space,
wiggly lines are associated with $\alpha(\tau_j-\tau^\prime_j)$.
Repeated blip-antiblip pairs (a), rainbows (b) and crossings (c)
are retained in for the TLS approximation. The diagram in (d)
involves four charge states.
\vspace{0.5cm}

\noindent
Fig. 4 :  The effective gap $\Delta$ and tunneling strength
$\alpha$ obtained by integrating the scaling equations for
different initial values
($\Delta_0 / E_C = \, 0.01\, , \; 0.05\, ,\; 0.1\,$),
compared with the values inferred from Monte Carlo simulations
(diamonds).
\vspace{0.5cm}

\noindent
Fig. 5 :  Monte Carlo results (diamonds) for the effective capacitance
renormalization in the weak tunneling regime and in the
(non-perturbative) intermediate tunneling regime. Comparison is made
with the results from perturbation theory, from Ref. \cite{kn:PanZai} and
from Ref. \cite{kn:GuiSch,kn:Kosterlitz}.
\vspace{0.5cm}

\noindent
Fig. 6 : The renormalized energy bands close to
the edges for various $\alpha_0$ in the weak
coupling limit.
\vspace{0.5cm}

\noindent
Fig. 7 : The normalized voltage at the junction
$C \langle V \rangle/e = \langle Q \rangle/e$
close to the band edges, at $T=0 \,$, for
various $\alpha_0$ in the weak coupling limit.
\vspace{0.5cm}

\noindent
Fig. 8 : The expectation value of the number of
excess electrons in the island close to the band
edges, at $T=0 \,$, for various $\alpha_0$ in the
weak coupling limit. In the absence of charging
effects the ``ohmic'' linear dependence is found.
\vspace{0.5cm}

\noindent
Fig. 9 :  Flow diagram in the
$\,\,\frac{E_J}{E_C}\,$ -- $\,\alpha\,\,$
plane for S-S tunnel junctions with finite subgap
conductance $\alpha_0$.

\end{document}